\documentclass[11pt]{article}
\usepackage{graphicx}
\usepackage{feynmf}
\usepackage{epsfig}
\usepackage{here}
\usepackage{wrapfig}
\usepackage{psfrag}
\usepackage{times}
\usepackage{lscape}
\usepackage{rotating}

\newcommand{\BABARPubYear}    {02}

\newcommand{\BABARConfNumber} {016}
\newcommand{\SLACPubNumber} {9297}

\def\Bflav {\ensuremath{B_{flav}}}
\def\Btag {\ensuremath{B_{tag}}}

\input babarsym.tex

\long\def\inst#1{\par\nobreak\kern 4pt\nobreak
    {\it #1}\par\vskip 10pt plus 3pt minus 3pt}

\begin{document}
{\pagestyle{empty}

\begin{flushright}
\babar-CONF-\BABARPubYear/\BABARConfNumber \\
SLAC-PUB-\SLACPubNumber \\
July 2002 \\
\end{flushright}

\begin{center}
\end{center}

\par\vskip 5cm

\begin{center}
\Large \bf Measurement of \boldmath{$\sin 2\beta$} in \boldmath{$B^0\to \phi \KS$}
\end{center}
\bigskip

\begin{center}
\large The \babar\ Collaboration\\
\mbox{ }\\
July 24, 2002
\end{center}
\bigskip \bigskip

\begin{center}
\large \bf Abstract
\end{center}

\begin{center}
\parbox{13.8cm}{
We present a preliminary measurement of the time-dependent $CP$-violating 
asymmetry in the neutral B decay 
$B^0\to \phi \KS$, with $\phi \to K^+K^-$ and $\KS \to \pi^+\pi^-$. 
The measurement uses a data sample of about 87 million $\FourS\to B\Bbar$
decays collected between 1999 and 2002 with the \babar\ detector at the
\pep2\ asymmetric-energy \BF\ at SLAC. In this sample we study events 
in which the \CP final state is fully reconstructed and the flavor of the 
other neutral $B$ meson is determined from its decay products. 
The amplitude of the \CP-violating asymmetry \stwob 
is derived from the decay-time distributions. We measure 
$\stwob = -0.19 ^{+0.52}_{-0.50} \stat \pm 0.09 \syst$.
}  
\end{center}

\vfill
\begin{center}
Contributed to the 31$^{st}$ International Conference on High Energy Physics,\\ 
7/24---7/31/2002, Amsterdam, The Netherlands
\end{center}

\vspace{1.0cm}
\begin{center}
{\em Stanford Linear Accelerator Center, Stanford University, 
Stanford, CA 94309} \\ \vspace{0.1cm}\hrule\vspace{0.1cm}
Work supported in part by Department of Energy contract DE-AC03-76SF00515.
\end{center}

\newpage
} 

\begin{center}
\small

The \babar\ Collaboration,
\bigskip

B.~Aubert,
D.~Boutigny,
J.-M.~Gaillard,
A.~Hicheur,
Y.~Karyotakis,
J.~P.~Lees,
P.~Robbe,
V.~Tisserand,
A.~Zghiche
\inst{Laboratoire de Physique des Particules, F-74941 Annecy-le-Vieux, France }
A.~Palano,
A.~Pompili
\inst{Universit\`a di Bari, Dipartimento di Fisica and INFN, I-70126 Bari, Italy }
% ...

\end{center}\newpage

\section{Introduction}
\label{sec:Introduction}
Recent measurements of the $CP$-violating asymmetry parameter
\stwob by the \babar~\cite{sin2bold} and Belle~\cite{sin2bbelle} 
collaborators established $CP$ violation in the $B^0$ system.
These measurements, as well as the updated measurement 
of $\sin2\beta = 0.741\pm 0.067(stat)\pm 0.033(syst)$ by
\babar~\cite{sin2bnew} reported at this conference, are consistent
with the Standard Model expectation based on measurements and
theoretical estimates of the Cabbibo-Kobayashi-Maskawa 
quark-mixing matrix~\cite{ckm}.

\begin{center}
\begin{figure}[htbp]
\setlength{\epsfxsize}{1.0\linewidth}\leavevmode\epsfbox{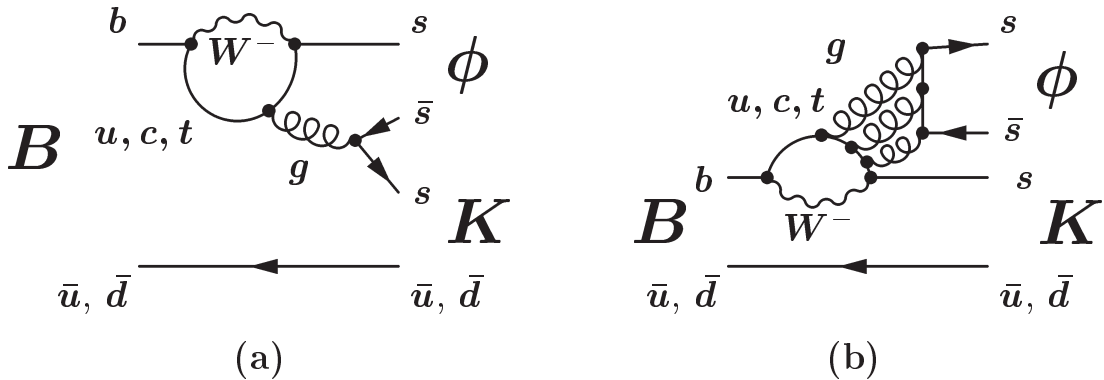}
\caption{Quark-level diagrams describing the decays \boldmath{$B\to\phi K$}: 
(a) internal penguin, (b) flavor-singlet penguin.}  
\label{fig:diagram} 
\end{figure}
\end{center}

Charmless hadronic $B$ meson decays provide important 
information for the study of $CP$ violation effects.
The charmless $B$ meson decays into final states with a $\phi$ 
meson are interesting because they are dominated by 
$b\to s\bar{s}s$ gluonic penguins (Figure~\ref{fig:diagram}),
with a smaller contribution from electroweak penguins,
while other Standard Model contributions are highly 
suppressed. These decays allow the extraction of the 
$CP$-violating parameter \stwob .
Comparison of the value of \stwob\ obtained from these modes
with that from charmonium modes probe for new physics
participating in penguin loops~\cite{grossman,fleischer}.
The predicted deviation of the effective 
$\stwob$ for the $\phi \KS$ mode from \stwob\
in the Standard Model is smaller than 4\%~\cite{grossman,grossman1}.
In this analysis we probe for sizable deviations which 
are possible in many scenarios beyond the Standard Model.
\vspace{-1cm}

The decay of neutral $B$ mesons to the $CP$~=~-1 final state $\phi \KS$
has been observed by \babar\ in a sample of about 45 million $B$ mesons 
with a branching fraction of 
$BF(B^0\to\phi K^0) = (8.1^{+3.1}_{-2.5}\pm0.8)\times 10^{-6}$~\cite{oldpub}.
The channel $B^+\to \phi K^+$, which is used as control channel
for the time-dependent analysis, was also observed with a branching
fraction of $(7.7^{+1.6}_{-1.4}\pm 0.8)\times 10^{-6}$~\cite{oldpub}.

The measurement of the time-dependent $CP$ asymmetry in $B^0\to \phi \KS$
is similar to our approach in the charmonium channels~\cite{sin2bprd}.
We use an extended parametrization of the likelihood describing
the event yield in signal and background which is combined with the 
likelihood for the decay-time distributions. 

\section{The \babar\ Detector and Data Set}
\label{sec:babar}
This measurement is based on data recorded with the \babar\ detector \cite{Aubert:2001tu} 
at the \pep2\ energy-asymmetric 
\epem storage ring at SLAC. The data sample corresponds to an integrated luminosity 
of approximately 80\invfb that was collected at the \FourS\ resonance.

The detector consists of a five-layer silicon vertex tracker (SVT), a 40-layer drift chamber
(DCH), a detector of internally reflected Cherenkov light (DIRC), an
electromagnetic calorimeter (EMC), assembled from 6580 CsI(Tl) crystals, all embedded in a 
solenoidal magnetic field of 1.5 T and surrounded by an instrumented flux return (IFR). 
The performance of the detector is discussed in \cite{sin2bprd}.

\section{Analysis Method}
\label{sec:Analysis}

\subsection{Time dependent Analysis}
Each candidate event consists of a fully reconstructed neutral $B$ meson, $B_{CP}$,
decaying into $\phi \KS$ and a partially reconstructed recoil $B$, $B_{tag}$, which
we examine for evidence that it decayed as \Bz or \Bzb (flavor tag). 
The decay-time distribution of $B$ decays to a \CP eigenstate with a \Bz
or \Bzb tag can be expressed in terms of a complex parameter $\lambda$
that depends on both the \Bz-\Bzb oscillation amplitude and the amplitudes
describing \Bzb and \Bz decays to this final
state~\cite{lambda}. The decay rate  ${\rm f}_+({\rm f}_-)$ when the 
tagging meson is a $\Bz (\Bzb)$ is given by 

\begin{eqnarray}
{\rm f}_\pm(\, \deltat) = {\frac{{\rm e}^{{- \left| \deltat \right|}/\tau_{\Bz} }}{4\tau_{\Bz}
}}  \times  \left[ \ 1 \hbox to 0cm{}
\pm \frac{{2\mathop{\cal I\mkern -2.0mu\mit m}}
\lambda}{1+|\lambda|^2}  \sin{( \deltamd  \deltat )} 
\mp { \frac{1  - |\lambda|^2 } {1+|\lambda|^2} }  
  \cos{( \deltamd  \deltat) }   \right],
\label{eq:timedist}
\end{eqnarray}

\vskip12pt\noindent
where $\Delta t = t_{CP} - t_{tag}$ is the difference between 
the proper decay time of the reconstructed $B$ meson ($B_{CP}$) and 
the proper decay time of the tagging $B$ meson ($B_{tag}$),
$\tau_{\Bz}$ is the \Bz lifetime, and \deltamd is the 
\Bz-\Bzb oscillation frequency.
The sine term in Eq.~(\ref{eq:timedist}) is due to the interference between direct
decay and decay after flavor change, and the cosine term is due to the
interference between two or more decay amplitudes with different weak
phases. Evidence for \CP violation can be observed as a difference
between the \deltat distributions of \Bz- and \Bzb-tagged events or as
an asymmetry with respect to $\deltat = 0$ for either flavor tag. 
\par
In the Standard Model and 
for the case that the decay proceeds purely via $b\to s\bar{s}s$
gluonic penguin transitions, $\lambda=\eta_f e^{-2i\beta}$, and
the angle $\beta$ of the Unitarity Triangle of the three-generation 
CKM matrix~\cite{ckm1} is given as 
$\beta = arg \left [ -V_{cd} V_{cb}^\star / V_{td} V_{tb}^\star\right]$. 

Thus, the time-dependent $CP$-violating asymmetry is
\begin{eqnarray}
A_{\CP}(\deltat) &\equiv&  \frac{ {\rm f}_+(\deltat)  -  {\rm f}_-(\deltat) }
{ {\rm f}_+(\deltat) + {\rm f}_-(\deltat) } = -\eta_f \stwob
\sin{ (\deltamd \, \deltat )} ,  
\label{eq:asymmetry}
\end{eqnarray}

\vskip12pt\noindent
with $CP$ eigenvalue $\eta_f=-1$ for $\phi\KS$, and $\KS \to \pi^+\pi^-$.

\subsection{Reconstruction of the \boldmath{$\phi\KS$} Final State}

We fully reconstruct $B$ meson candidates ($B_{CP}$) in the decay mode
$\phi \KS$ with $K^0_S\rightarrow\pi^+\pi^-$ and $\phi\rightarrow K^+K^-$.
For the charged tracks belonging to the recoil $B$ and to the
$\KS$ we require at least 12 measured drift chamber hits and a minimum
transverse momentum of 0.1~GeV/$c$.
The kaon tracks of the $\phi$ in addition have to originate within
1.5~cm in the transverse 
plane and 10~cm in beam direction
from the interaction point.

A track is identified as a kaon 
based on a likelihood ratio combining the $dE/dx$
information from the SVT and DCH below 700~MeV/$c$, and from DCH $dE/dx$ and 
DIRC Cherenkov angle and measured Cherenkov photon number
above this momentum. 
We define $\phi$ candidates as pairs of tracks with opposite 
charge which can be combined and fit to a common vertex and 
whose invariant $K^+K^-$ mass lies within a 20~MeV/$c^2$ mass interval 
centered at the $\phi$ mass.
In case there is no particle identification (PID) information,
the kaon hypothesis is assumed for one of the two tracks from
the candidate $\phi$ decay. 
Using a relativistic Breit-Wigner function of fixed mass and
width~\cite{pdg} convoluted with a Gaussian we obtain an invariant
mass resolution of 1.1~MeV/$c^2$ in the $\phi$ signal.

Analogously to the $\phi$, we construct the $\KS$ from two oppositely
charged tracks which are assumed to be pions. 
The selection of $K^0_S\rightarrow\pi^+\pi^-$ candidates is based
on the angle $\alpha$ between the line connecting $\phi$ vertex and
$\KS$-decay vertex and the momentum direction reconstructed from 
the pions ($\cos\alpha > 0.999$). Furthermore, we use the 
decay-time significance $t/\sigma_t$ ($t/\sigma_t > 3$).

\subsection{Event Yield Variables}
The measurement of $B$ decays at the $\Upsilon(4S)$
resonance provides kinematic constraints for the initial state.
Substitution of the measured energy by the beam energy reduces 
the resolution of kinematic variables substantially. 

Energy resolution can be expressed as:
\begin{equation}
\Delta E = E_B - E_{bc} \, ,
\end{equation}
with $E_{bc}$ the beam constrained energy, which for the candidate
$B$ meson is derived as follows:
\begin{equation}
E_{bc} = \frac{s + 2 \vec{p}_i \cdot \vec{p}_B}{2 E_i}\, ,
\end{equation}
with $\sqrt{s}$ the total $e^+e^-$ center-of-mass energy.
The four momentum of the initial state is represented by $(E_i,\vec{p}_i)$, 
and $(E_B,\vec{p}_B)$ is the four momentum of the candidate $B$ meson,
both measured in the laboratory;
$E_{bc}$ results from the assumption that we have particle-antiparticle
production. Notice that the $B$-candidate momentum $\vec{p}_B$
is independent of the mass values assigned
to the tracks comprising the candidate $B$.
Signal events distribute in $\Delta E$ according to a Gaussian with a 
mean consistent with zero ($\pm 2$~MeV). The observed width is about $17$~MeV.
The background shape in $\Delta E$ is parametrized by a linear function.
We require $|\Delta E| < 200$~MeV.

The second  kinematic quantity in our analysis is 
the beam-energy substituted mass $m_{ES}$, which is defined as:
\begin{equation}
 m_{ES} = \sqrt{ E_{bc}^2 - \vec{p}_B^{\,2} }.
\end{equation}
Signal events are distributed Gaussian-like in $m_{ES}$ with
a mean at the $B$ mass and a resolution of about 2.6~MeV/$c^2$,
dominated by the beam energy spread. 
The background shape in $m_{ES}$ is parametrized by a threshold (ARGUS)
function~\cite{argus} with a fixed endpoint given by the average beam
energy. Our selection requires $m_{ES} > 5.22$~GeV/$c^2$.

The helicity angle $\theta_H$ of the $\phi$ is defined as the
angle between the direction of the decay $K^+$ and the parent $B$
direction in the $\phi$ rest frame. For pseudoscalar-vector
$B$ decay modes, angular momentum conservation results in a
$\cos^2\theta_H$ distribution. In this variable the background is 
uniformly distributed. 

Monte Carlo simulation demonstrates that contamination from
other $B$ decays is negligible. 
Possible ($K^+K^-$) S-wave contributions ($f_0(980)$) are not
expected to peak under the $\phi$ meson~\cite{scalar} and are
suppressed by the helicity angle which distributes uniformly
for this background.
However, charmless hadronic
modes suffer from backgrounds due to random combinations of tracks 
produced in the quark-antiquark ($\bar{q}q$) continuum,
where $q$ is dominantly $u$, $d$, and $s$ quarks.
The distinguishing feature of such backgrounds is their
characteristic event shape resulting from the two-jet production
mechanism.

We consider the angle $\theta_T$ between the thrust axis of
the $B$ candidate and the thrust axis of the rest of the
event, where the thrust axis is defined as the axis that 
maximizes the sum of the magnitudes of the longitudinal
momenta in the $\Upsilon(4S)$ center-of-mass system.
This angle is small for continuum events where
the $B$-candidate daugthers come from back-to-back $\bar{q}q$ jets,
and is uniformly distributed for true $B\bar{B}$ events.
In the event preselection we require $|\cos\theta_T| < 0.9$.

Additional shape information comes from the momentum flow around the $B$ thrust axis
described through momentum weighted Legendre polynomials, $L_i$. The best separation between
the signal and continuum events is achieved with the zeroth order ($L_0=\sum p_i^\star$)
and the second order ($L_2=\sum p_i^\star \times \frac{1}{2}(\cos^2{\theta_i^\star}-1) $)  
polynomials, where $p_i^\star$ and $\theta_i^\star$ are the center-of-mass
momentum and angle with respect to the $B_{CP}$ thrust axis and
the sum is over all charged tracks and neutrals in the event that are not 
associated with the $B$-candidate.

The last event shape variable is the $B$ production angle, $\theta_B$, with respect to the beam 
direction in the \FourS\ center-of-mass frame. In decays of a real \FourS\ into two pseudoscalar $B$ mesons, 
the production angle follows a $\sin^2 \theta_B$ distribution, while it is approximately uniformly
distributed for the continuum events.

The shape variables are strongly correlated and cannot be used independently in the likelihood calculation.
A Fisher discriminant is formed as a linear combination of the shape variables
$x=|\cos \theta_T|$, $\cos \theta_B$, $L_0$, and $L_2$:
\begin{equation}
	{\mathcal F} = \sum \gamma_i x_i ,
\end{equation}
where coefficients $\gamma_i$ are chosen such to make the maximum separation between
the signal and continuum event distributions.
The coefficients are calculated using Monte Carlo signal events and background
events from data sidebands ($0.1<\Delta E<0.3$).
For the resulting signal Fisher distribution we use a bifurcated Gaussian distribution.
The background Fisher shape is described by a sum of two Gaussian distributions.   

\section{Tagging and Vertexing}

We use a $B$-tagging algorithm based on multivariate techniques
to determine the flavor of $B_{tag}$~\cite{sin2bnew}.  
The algorithm relies on the correlation between the flavor of the $b$ quark and the charge 
of the remaining tracks in the event after removal of the $B_{CP}$ candidate.  
Separate neural networks are trained to identify primary leptons from semileptonic
$B$ decay, kaons, soft pions from $D^*$ decay, and high-momentum charged particles.
The outputs of each neural network are combined to produce five hierachical and
mutually exclusive tagging categories.
Events with an identified electron or muon, 
and a supporting kaon, if present
are assigned to the {\tt Lepton} category. 
Events with an identified kaon and a soft-pion candidate with opposite charges are 
assigned to the {\tt Kaon\,I} category.  Events with one or more kaon candidates,
and no lepton or soft-pion candidates, are assigned to the {\tt Kaon\,I} or {\tt Kaon\,II}
categories depending on the 
estimated mistag probability.
Events with only 
a soft-pion candidate are assigned to the {\tt Kaon\,II} category
as well.  
The remaining events
are assigned to the {\tt Inclusive} 
or {\tt Untagged} category
based on estimated mistag probability.

The quality of tagging is expressed in terms of the effective efficiency 
$Q = \sum_c \epsilon_c (1-2w_c)^2$, where $\epsilon_c$ and $w_c$ are the 
efficiency and mistag probability, respectively, for events tagged in category $c$.
Table~\ref{tab:tagging} summarizes the tagging performance in a data sample of 
fully reconstructed neutral $B$ decays into $D^{(*)-}h^+\,(h^+ = \pip, \rho^+, a_1^+)$ 
and $\jpsi K^{*0}\,(K^{*0}\to\Kp\pim)$ flavor eigenstates (\Bflav\ sample).
The recoil $B$, $B_{tag}$, is again partially reconstructed.
We use the same tagging efficiencies and dilutions for the $\phi \KS$ channel 
extracted from the statistical dominant flavor sample.

\begin{table}[!tbp]
\begin{center}
\parbox{14cm}{
\caption{Tagging efficiency $\epsilon$, average mistag fraction $w$, 
mistag difference $\Delta w = w(\Bz) - w(\Bzb)$, and effective tagging 
efficiency $Q$ for signal events in each tagging category.  The values are 
measured in the \Bflav\ sample.} 
\label{tab:tagging} }
\vskip 4mm
\begin{tabular}{lrrrr} \hline\hline
Category & $\epsilon\,(\%)$ & $w\,(\%)$ & $\Delta w\,(\%)$ & $Q\,(\%)$ \rule[-2mm]{0mm}{6mm} \\\hline
{\tt Lepton}    & $ 9.1 \pm 0.2$ &  $3.3 \pm 0.6$ & $-1.5 \pm 1.1$ &  $7.9 \pm 0.3$ \\
{\tt Kaon\,I}    & $16.7 \pm 0.2$ & $10.0 \pm 0.7$ & $-1.3 \pm 1.1$ & $10.7 \pm 0.4$ \\
{\tt Kaon\,II}     & $19.8 \pm 0.3$ & $20.9 \pm 0.8$ & $-4.4 \pm 1.2$ &  $6.7 \pm 0.4$ \\
{\tt Inclusive} & $20.0 \pm 0.3$ & $31.5 \pm 0.9$ & $-2.4 \pm 1.3$ &  $2.7 \pm 0.3$ \\
{\tt Untagged}  & $34.4 \pm 0.5$ &                &                &                \\\hline
Total $Q$       &                &                &                & $28.1\pm 0.7$ \rule[-2mm]{0mm}{6mm} \\\hline\hline
\end{tabular}
\end{center}
\end{table}

The time difference $\deltat$ is obtained from the measured distance between 
the $z$ positions of the $B_{CP}$ and $B_{tag}$ decay vertices and the known boost 
of the $\epem$ system.  
For the $B_{CP}$ we achieve a $z$-vertex position resolution of better than 60~$\mu m$ 
which compares well to the resolution obtained in the final state $J/\Psi \KS$.
The $z$ position of the \Btag\ vertex is determined 
with an iterative procedure that removes tracks with a large contribution to 
the total $\chi^2$.  An additional constraint is constructed from the three-momentum 
and vertex position of the $B_{\rm CP}$ candidate, and the average $\epem$ interaction 
point and boost.  For $98\%$ of candidates with a reconstructed vertex the 
r.m.s.\ $\deltaz$ resolution is $180\mum\,(1.1\ps)$.  
We require $\left|\deltat\right|<20\ps$ and $\sigma_{\deltat} < 3.5\ps$, where 
$\sigma_{\deltat}$ is the event-by-event error on $\deltat$.

The $\Delta t$ resolution is dominated by the tag-side ($B_{tag}$),
which is well under control from our flavor sample.
The empirical $\Delta t$ resolution function for signal candidates is
parametrized by a sum of three Gaussian distributions~(see Ref.~\cite{BaBarSin2betaM02}), 
with parameters determined from a fit to the flavor sample. 
A common parametrization is used for all tagging categories, 
and the parameters are determined simultaneously with the \CP parameters in the maximum 
likelihood fit. The tagging parameters for the untagged events are fixed to $w = 0.5$
and $\Delta w = 0$.
The $\deltat$ background in the $\phi\KS$ channel does not show 
a lifetime component and is parametrized as a sum of two Gaussians, core and tail,
with the tail fraction of 2\%. 

The parametrization of $\deltat$ 
is checked by measuring the lifetime in the channel $\phi K^+$ (180 signal events)
which has a compatible $\deltat$ distribution.
The obtained value agrees within 1$\sigma$ with the world average~\cite{pdg}.

\subsection{Maximum Likelihood Fit}

We use an unbinned extended maximum likelihood fit to extract yields and $\CP$ parameters
from the $\phi \KS$ (CP) sample, and simultaneously $\deltat$ resolution parameters
and tagging quantities in the flavor sample.
The likelihood for candidate $j$ tagged in category 
$c$ is obtained by summing the product of event yield $n_{i}$, tagging efficiency $\epsilon_{i,c}$,
and probability ${\cal P}_{i,c}$ over the two possible signal and background hypotheses $i$.

\begin{equation}
{\cal L}_c = \frac{1}{N!}\exp{\left(-\sum_{i}n_i\epsilon_{i,c}\right)}
\prod_{j}\left[\sum_{i}n_i\epsilon_{i,c}{\cal P}_{i,c}(\vec{x}_j;\vec{\alpha}_i)\right].
\end{equation}
The probabilities ${\cal P}_{i,c}$ are evaluated as the product of PDFs 
for each of the independent variables 
$\vec{x}_j = \left\{\mes, \Delta E, {\cal F}, \cos\theta_H, \deltat\right\}$
in the $CP$ sample, and $\vec{x}_j = \left\{\mes, \Delta E, \deltat\right\}$ 
in the $\Bflav$ sample. The $\vec{\alpha}_i$ are fixed parameters that describe
the expected distributions and are derived from fits to signal Monte Carlo,
the $B^+\to\phi K^+$ control channel, on-resonance sidebands, and off-resonance 
data. The \deltat distributions of the $B_{\rm flav}$ sample evolve
according to flavor oscillation in $B^0$
mesons. The observed amplitudes for the \CP asymmetry in the
$CP$ sample and for flavor oscillation in the flavor sample 
are reduced by the same factor $1-2\mistag$ due to flavor mistags. 
The total likelihood ${\cal L}$ is the product of likelihoods for each tagging 
category and the free parameters are determined by minimizing the quantity 
$-\ln{\cal L}$~\cite{minuit}.

The total number of $\phi\KS$ candidate events in the fit region is 1352.
In order to extract the event yields we perform an initial fit without tagging or $\deltat$ 
information. Our final $CP$ sample is composed of 51 signal and 1301 background events 
distributed over the fit range. 
For our $CP$ fit we fix these yields. The $\Bflav$ sample consists of about 26000 
signal events with a purity of better than 80\%.

There are 34 free variables in the fit, in agreement with the 
charmonium $\sin2\beta$ fit~\cite{sin2bnew}, 
where only the parameter $\sin 2\beta$
is solely fit from the signal in $B^0\to \phi\KS$ ($|\lambda| = 1$ fixed).
The other parameters are 
the average mistag fraction $\mistag$ and the 
difference $\Delta\mistag$ between \Bz\ and \Bzb\ mistags for each
tagging category (8), parameters for the signal \deltat resolution (8), 
parameters for background time dependence (6), background \deltat resolution
(3), and mistag fractions (8). 
The determination of the mistag fractions and \deltat resolution
function parameters for the signal is dominated by the high-statistics 
flavor sample. 
We fix $\tau_{\Bz}=1.542\ps$ and $\deltamd =0.489\ps^{-1}$~\cite{pdg}. 
The largest correlation between \stwob\ and any linear combination of the other 
free parameters is 2\%. 

The result for the effective $\sin2\beta$ is:
\begin{equation}
\sin2\beta  = -0.19 ^{+0.52}_{-0.50} (stat) \pm 0.09  (syst)
\end{equation}

\unitlength1.0cm 
\begin{figure}[ht]
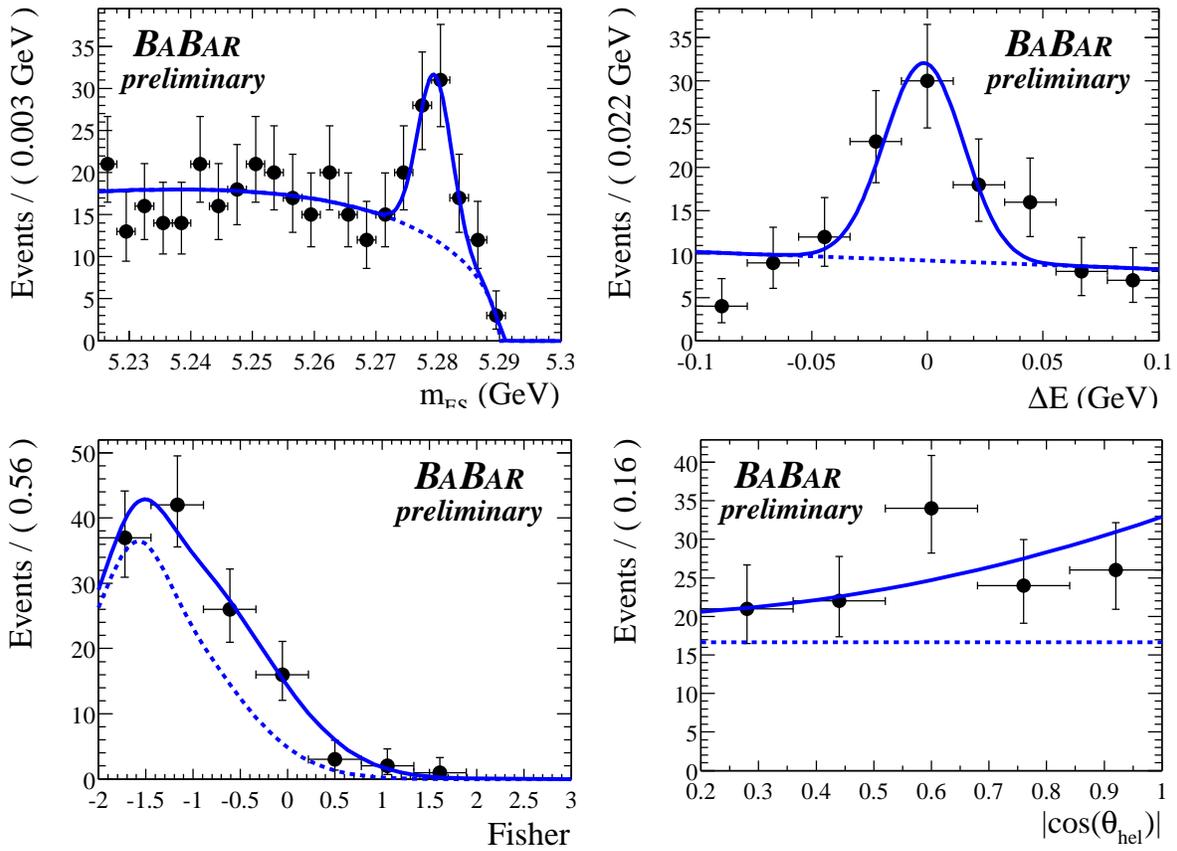

\center
\begin{tabular}{cc}
\epsfig{file=pdfProjectionsForAllEvents_mES.epsi, width=5.4cm, angle=270} &
\epsfig{file=pdfProjectionsForAllEvents_deltaE.epsi, width=5.4cm, angle=270} \\
\epsfig{file=pdfProjectionsForAllEvents_F.epsi, width=5.4cm, angle=270} &
\epsfig{file=pdfProjectionsForAllEvents_cosHel.epsi, width=5.4cm, angle=270} 
\end{tabular}
\caption{The distribution of the $\phi \KS$ events in four fit variables
with tighter selection criteria corresponding to the ranges shown
(see text). The $m_{ES}$ distribution is presented in a wider range. 
The solid line refers to the fit 
for all events, the dashed line corresponds to the expected background 
distribution.}
\label{fig:yield}
\end{figure}

\begin{figure}[h]
\center
\parbox{14cm}{ 
$B^0$ tags: \\

\epsfig{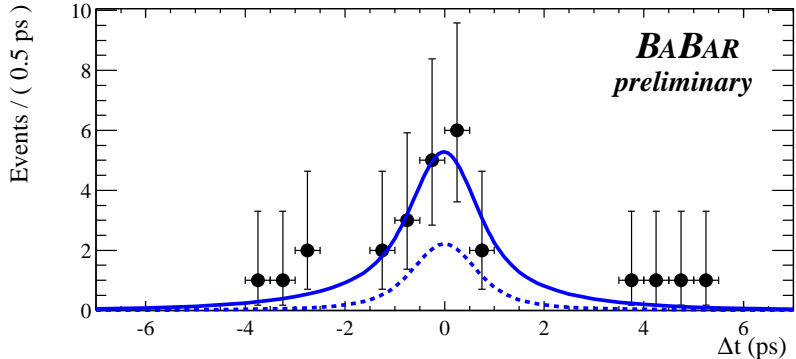} }
\parbox{14cm}{ 
$\bar{B}^0$ tags: \\

\epsfig{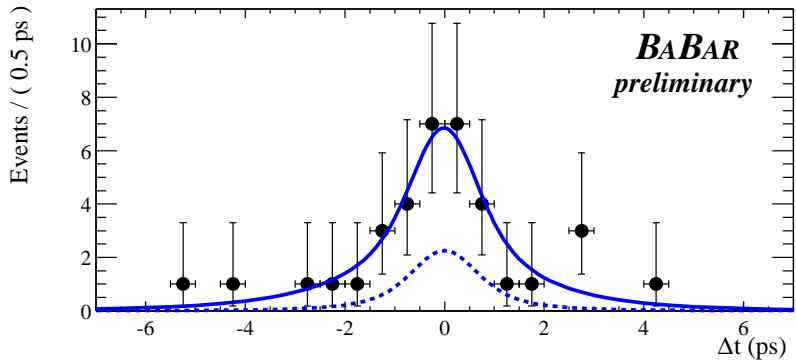} }
\caption{Time distributions for  $B^0$ and  $\bar{B}^0$ tags. The solid line refers to 
the fit for all events, the dashed line corresponds to background.
\label{fig:b0b0bar} }
\end{figure}

Figure~\ref{fig:yield} shows the event yield variables for all events
in the limited ranges $5.27 < m_{ES} < 5.3$~GeV/$c^2$, $|\Delta E| < 0.1$~GeV,
$-2 < \mathcal{F} < 3$, and $|\cos\theta_H| > 0.2$, focussing into the
signal region. Figure~\ref{fig:b0b0bar} shows the $\deltat$ 
distributions for the $B^0$ and the $\bar{B}^0$ tagged subsets of
these events with the fit superimposed.

As a consistency check we measure the \CP asymmetry in the
final state $B^+\to \phi K^+$. From a sample of 180 signal events
we obtain a \CP asymmetry of  $\sin 2\beta(\phi K^+) = 0.26\pm 0.27$,
which is consistent with the expected value of zero.

Repeating the fit with all parameters except \stwob
fixed to their values at the maximum likelihood, we attribute a
total contribution in quadrature of $0.01$ to the
error on \stwob\ due to the combined statistical uncertainties in
mistag rates, \deltat\ resolution, and background parameters. 

The fit is repeated on generated datasets based on the 
probability density functions for signal and background shapes. 
We do not observe a bias in the refit value of $\sin2\beta$.

We consider systematic uncertainties due to 
the event yield determination in $\phi \KS$ (0.02),
limited Monte Carlo statistics (0.02), composition and
\CP asymmetry in the background in the \CP events (0.03),
the assumed parametrization of the $\Delta t$ resolution function
(0.02), due in part to residual uncertainties in the
Silicon Vertex Tracker alignment, and the
fixed values for $\Delta m_d$ and $\tau_B$ (0.006).

Furthermore, we explore the sensitivity to the parameter $|\lambda|$
in our limited sample. It turns out that the fit is not simultaneously 
sensitive to both $\sin 2\beta$ and $|\lambda|$.
Therefore, we scan for the value of $|\lambda|$ over a wide range
($|\lambda| = 0 ... 3$) but do not observe a strong variation of the 
central value of $\sin2\beta$. We attribute an additional conservative 
error of 0.08 to our value of $\sin 2\beta$, which is the maximum variation
observed in the scan.

\section{Conclusions}
\label{sec:Physics}

We measure the preliminary effective value of the time-dependent 
\CP asymmetry $\sin2\beta = -0.19 ^{+0.52}_{-0.50} (stat)$ $\pm 0.09 (syst)$
in the decay of neutral $B^0_d$ mesons into the final state
$\phi \KS$, $\KS\to\pi^+\pi^-$.
The deviation of this value from the updated \babar\ value
presented at this conference,
$\sin2\beta = 0.741 \pm 0.067 (stat) \pm 0.033 (syst)$,
is about two standard deviations.
The measurement will, for some time, be dominated by the statistical 
uncertainty.

\section{Acknowledgments}
\label{sec:Acknowledgments}

We are grateful for the 
extraordinary contributions of our \pep2\ colleagues in
achieving the excellent luminosity and machine conditions
that have made this work possible.
The success of this project also relies critically on the 
expertise and dedication of the computing organizations that 
support \babar.
The collaborating institutions wish to thank 
SLAC for its support and the kind hospitality extended to them. 
This work is supported by the
US Department of Energy
and National Science Foundation, the
Natural Sciences and Engineering Research Council (Canada),
Institute of High Energy Physics (China), the
Commissariat \`a l'Energie Atomique and
Institut National de Physique Nucl\'eaire et de Physique des Particules
(France), the
Bundesministerium f\"ur Bildung und Forschung and
Deutsche Forschungsgemeinschaft
(Germany), the
Istituto Nazionale di Fisica Nucleare (Italy),
the Research Council of Norway, the
Ministry of Science and Technology of the Russian Federation, and the
Particle Physics and Astronomy Research Council (United Kingdom). 
Individuals have received support from 
the A. P. Sloan Foundation, 
the Research Corporation,
and the Alexander von Humboldt Foundation.

\bibliographystyle{h-physrev2-original}   %

\end{document}